\documentclass[iop,apjl,twocolumn,single]{emulateapj}
\usepackage{enumitem,graphicx,apjfonts,multirow,amsmath,bm}

\shorttitle{Cyclic Convective Dynamo}
\shortauthors{Augustson et al.}

\newcommand{\Rsun}[1]{#1\,\mathrm{R} \,\!\scriptscriptstyle \sun\!}
\newcommand{\Osun}[1]{#1\,\Omega \,\!\scriptscriptstyle \sun\!}
\newcommand{\Prt}{\mathrm{Pr}}
\newcommand{\Prm}{\mathrm{Pm}}

\newcommand{\grad}{\boldsymbol{\nabla}}

\newcommand{\cross}{\boldsymbol{\times}}
\newcommand{\cnabla}{\boldsymbol{\cdot}\boldsymbol{\nabla}}

\newcommand{\vB}{\mathbf{B}}
\newcommand{\vJ}{\mathbf{J}}

\newcommand{\Bcp}{B_{\phi}}
\newcommand{\Lp}{L_{\phi}}

\newcommand{\avg}[1]{\langle #1 \rangle}
\newcommand{\avgO}{\langle\Omega\rangle}
\newcommand{\avgBp}{\langle B_{\phi} \rangle}

\begin{document}

\title{Cycling Dynamo in a Young Sun: Grand Minima and Equatorward Propagation}
\author{Kyle\ Augustson$^1$, Sacha\ Brun$^2$, Mark\ Miesch$^3$, \& Juri\ Toomre$^{1}$}
\affil{$^{1}$JILA and Dept. of Astrophysical \& Planetary Sciences University of Colorado, Boulder, CO 80309, USA}
\affil{$^{2}$DSM/IRFU/SAp, UMR AIM, Paris-Saclay CEA Saclay, 91191 Gif-sur-Yvette Cedex, France}
\affil{$^{3}$High Altitude Observatory, Center Green 1, Boulder, CO 80301, USA}
\email{Kyle.Augustson@colorado.edu}

\begin{abstract}

  We assess the global-scale dynamo action achieved in a simulation of a sun-like star rotating at
  three times the solar rate. The 3-D MHD Anelastic Spherical Harmonic code, using slope-limited
  diffusion, is employed to capture convection and dynamo processes in such a young sun. The
  simulation is carried out in a spherical shell that encompasses 3.8 density scale heights of the
  solar convection zone. We find that dynamo action with a high degree of time variation occurs,
  with many periodic polarity reversals every 6.2~years. The magnetic energy also rises and falls
  with a regular period, with two magnetic energy cycles required to complete a polarity
  cycle. These magnetic energy cycles arise from a Lorentz-force feedback on the differential
  rotation, whereas the polarity reversals are present due to the spatial separation of the
  equatorial and polar dynamos. Moreover, an equatorial migration of toroidal field is found, which
  is linked to the changing differential rotation and to a near-surface shear layer. This simulation
  also enters a grand minimum lasting roughly 20~years, after which the dynamo recovers its regular
  polarity cycles.

\end{abstract}

\keywords{stars: late-type, magnetic field, rotation -- convection -- magnetohydrodynamics -- turbulence}

\section{Introduction} \label{sec:intro}

The Sun exhibits many time scales from the ten minute lifetimes of granules to multi-millennial
magnetic activity modulations. One of the most prominent of these scales is the 11-year sunspot
cycle, during which the number of magnetically active regions waxes and wanes. The Sun also
possesses longer-term variability of its magnetic activity such as the 88-year Gleissberg cycle
\citep{gleissberg39} and less frequent phenomenon commonly described as grand extrema
\citep{usoskin13}. Other main-sequence stars also exhibit cyclical magnetic phenomenon in Ca II,
photometric, spectropolarimetric, and X-ray observations \citep[e.g.,][]{baliunas96, hempelmann96,
  favata08, metcalf10, fares13, mathur13}. These observations include solar-mass stars younger than
the Sun that also possess magnetic activity cycles, yet they rotate more rapidly than the Sun as a
consequence of the low rate of angular momentum loss in such stars \citep{barnes07}. Furthermore,
there are hints from both observations and from theory that a star's magnetic cycle period is
closely linked to its rotation rate \citep[e.g.,][]{saar09,jouve10,morgenthaler11}. This may imply
that the dynamo regime achieved in our simulation of a young sun, which rotates three times faster
than the Sun and has a nearly constant magnetic polarity cycle of 6.2~years, can scale up to the
solar rotation rate with a polarity cycle period closer to the 22~year cycle of the Sun.

In addition to its large range of time scales, observations of the magnetic field at the solar
surface reveal complex, hierarchical structures existing on a vast range of spatial scales. Despite
these chaotic complexities, large-scale organized spatial patterns such as Maunder's butterfly
diagram, Joy's law, and Hale's polarity laws suggest the existence of a structured large-scale
magnetic field within the solar convection zone. On the Sun's surface active regions initially
emerge at mid-latitudes and appear at increasingly lower latitudes as the cycle progresses, thus
exhibiting equatorward migration. As the low-latitude field propagates toward the equator, the
diffuse field that is comprised of small-scale bipolar regions migrates toward the pole, with the
global-scale reversal of the polar magnetic field occurring near solar maximum
\citep[e.g.,][]{hathaway10,stenflo12}.

Consequently, the large-scale field must vary with the solar cycle, likely being sustained through
dynamo action deep in the solar interior.  It has been suspected for at least 60 years that the
crucial ingredients for the solar dynamo are the shear of the differential rotation and the helical
nature of the small-scale convective flows present in the solar convection zone
\citep[e.g.,][]{parker55, steenbeck69, parker77}. Yet even with the advancement to fully nonlinear
global-scale 3-D MHD simulations \citep[e.g.,][]{gilman83,glatzmaier85,brun04,browning06}, achieving
dynamo action that exhibits the basic properties of Sun's magnetism has been quite
challenging. Nonetheless, recent global-scale simulations of convective dynamos have begun to make
substantial contact with some of the properties of the solar dynamo through a wide variety of
numerical methods \citep[e.g.,][]{brown11,racine11, kapyla12, nelson13a}. It is within this vein of
modern global-scale modeling that we report on a global-scale 3D MHD convective dynamo simulation
utilizing the ASH code that possesses some features akin to those observed during solar cycles.

\begin{figure*}[t!]
   \begin{center}
   \includegraphics[width=\textwidth]{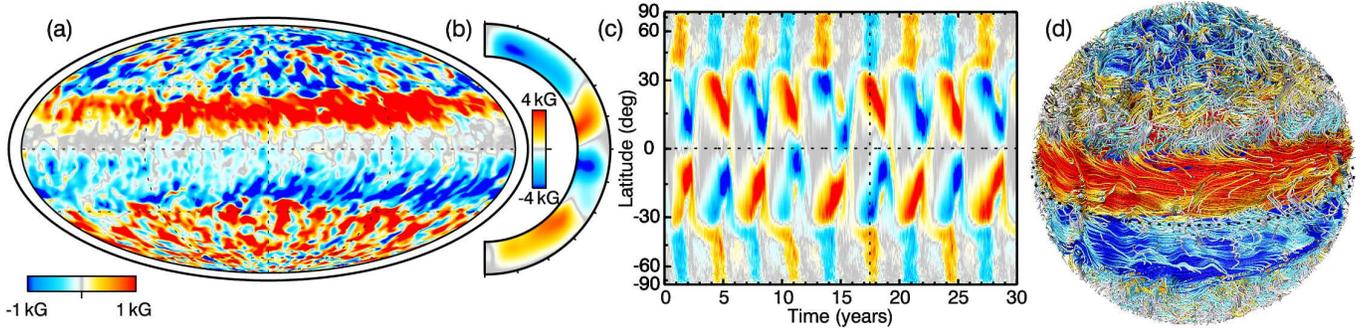}
   \figcaption{Nature of the toroidal magnetic field $\Bcp$. (a) Snapshot of the horizontal
     structure of $\Bcp$ at $\Rsun{0.95}$ shown in Mollweide projection, at the time corresponding
     to the vertical dashed line in (c). This illustrates the azimuthal connectivity of the magnetic
     wreaths, with the polarity of the field such that red (blue) tones indicate positive (negative)
     toroidal field. (b) Azimuthally-averaged $\avgBp$ also time-averaged over a single energy
     cycle, depicting the structure of the toroidal field in the meridional plane. (c) Time-latitude
     diagram of $\avgBp$ at $\Rsun{0.95}$ in cylindrical projection, exhibiting the equatorward
     migration of the wreaths from the tangent cylinder and the poleward propagation of the higher
     latitude field. The color is as in (a). (d) A rendering of magnetic field lines in the domain
     colored by the magnitude and sign of $\Bcp$, with strong positively oriented field in red, and
     the strong oppositely directed field in blue. \label{fig1}}
   \end{center}
\end{figure*}

\section{Methods} \label{sec:methods}

The 3D simulation of convective dynamo action presented here uses the ASH code to evolve the
anelastic MHD equations for a conductive calorically perfect plasma in a rotating spherical
shell. ASH solves the necessary equations with a spherical harmonic decomposition of the entropy,
magnetic field, pressure, and mass flux in the horizontal directions \citep{clune99,miesch00}. A
fourth-order non-uniform finite difference in the radial direction resolves the radial derivatives
\citep{featherstone13}. The solenoidality of the mass flux and magnetic vector fields is maintained
through the use of a stream function formalism \citep{brun04}. The boundary conditions used are
impenetrable on radial boundaries, with a constant entropy gradient there as well. The magnetic
boundary conditions are perfectly conducting at the lower boundary and extrapolated as a potential
field at the upper boundary. 

The authors have implemented a slope-limited diffusion (SLD) mechanism into the reformulated ASH
code, which is similar to the schemes presented in \citet{rempel09} and \citet{fan13}. SLD acts
locally to achieve a monotonic solution by limiting the slope in each coordinate direction of a
piecewise linear reconstruction of the unfiltered solution. The scheme minimizes the steepest
gradient, while the rate of diffusion is regulated by the local velocity. It is further reduced
through a function $\phi$ that depends on the eigth power of the ratio of the cell-edge difference
$\delta_i q$ and the cell-center difference $\Delta_i q$ in a given direction $i$ for the quantity
$q$. This limits the action of the diffusion to regions with large differences in the reconstructed
solutions at cell-edges. Since SLD is computed in physical space, it incurs the cost of smaller time
steps due to the convergence of the grid at the poles. The resulting diffusion fields are projected
back into spectral space and added to the solution.

We simulate the solar convection zone, stretching from the base of the convection zone at
$\Rsun{0.72}$ to the upper boundary of our simulation at $\Rsun{0.97}$. This approximation omits the
near-surface region and any regions below the convection zone. The SLD has been restricted to act
only on the velocity field in this simulation. This mimics a lower thermal and magnetic Prandtl
number ($\Prt$, $\Prm$) than otherwise attainable through an elliptic diffusion operator. The
entropy and magnetic fields remain under the influence of an anisotropic eddy diffusion, with both a
radially dependent entropy diffusion $\kappa_S$ and resistivity $\eta$. These two diffusion
coefficients are similar to those of case D3 from \citep{brown10}, with $\kappa_S , \eta \propto
\overline{\rho}^{\; -1/2}$, with $\overline{\rho}$ the spherically symmetric density. The
stratification in this case has about twice the density contrast across the domain, being 45 rather
than 26, and has a resolution of $N_r\times N_{\theta} \times N_{\phi} = 200\times256\times512$.

\section{Cyclical Convective Dynamo Action} \label{sec:cycles}

Global-scale convective dynamo simulations in rotating spherical shells have recently achieved the
long-sought goal of cyclical magnetic polarity reversals with a multi-decadal period. Moreover, some
of these simulations have illustrated that large-scale dynamo action is possible within the bulk of
the convection zone, even in the absence of a tachocline. Global-scale MHD simulations of a more
rapidly rotating Sun with the pseudo-spectral Anelastic Spherical Harmonic (ASH) code have produced
polarity reversing dynamo action that possesses strong toroidal wreaths of magnetism that propagate
poleward as a cycle progresses \citep{brown11}. These fields are seated deep within the convection,
with the bulk of the magnetic energy near the base of the convection zone. The perfectly conducting
lower boundary condition used here and in those simulations requires the field to be horizontal
there, which tends to promote the formation of longitudinal structure in the presence of a
differential rotation.

A recent simulation with ASH employs a dynamic Smagorinski diffusion scheme, wherefore they achieve
a greater level of turbulent complexity. Those simulations show that the large-scale toroidal
wreaths persist despite the greater pummeling they endure from the more complex and vigorous
convection \citep{nelson13a}. Not only do the toroids of field persevere, but portions of them can
be so amplified that the combination of upward advection and magnetic buoyancy create loops of
magnetic field \citep{nelson13b}. This lends credence to the classical picture of a Babcock-Leighton
or Parker interface dynamo \citep{leighton69,parker93}, with semi-buoyant flux structures that rise
toward the solar surface, leading to active regions and helicity ejection. There is the caveat that
the magnetic fields in the simulation are instead built in the convection zone.

Implicit large-eddy simulations (ILES) have concurrently paved the road toward more orderly
long-term cycles in a setting that mimics the solar interior. Indeed, simulations utilizing the
Eulerian-Lagrangian (EULAG) code produce regular polarity cycles occurring roughly every 80 years in
the presence of a tachocline and with the bulk of the magnetic field existing at higher latitudes
\citep{ghizaru10}. This simulation showed radial propagation of structures but little latitudinal
variation during a cycle. More recent simulations of a Sun-like star rotating at $\Osun{3}$ also
produce low-latitude poleward propagating solutions \citep{charbonneau13}. Such dynamo action is
accomplished first through the reduction of the enthalpy transport of the largest scales through a
simple sub-grid-scale (SGS) model that diminishes thermal perturbations over a roughly 1.5~year time
scale, which serves to moderate the global Rossby number. The ILES formulation of EULAG also
maximizes the complexity of the flows and magnetic fields for a given Eulerian grid resolution.

\begin{figure*}[t!]
   \begin{center}
     \includegraphics[width=\textwidth]{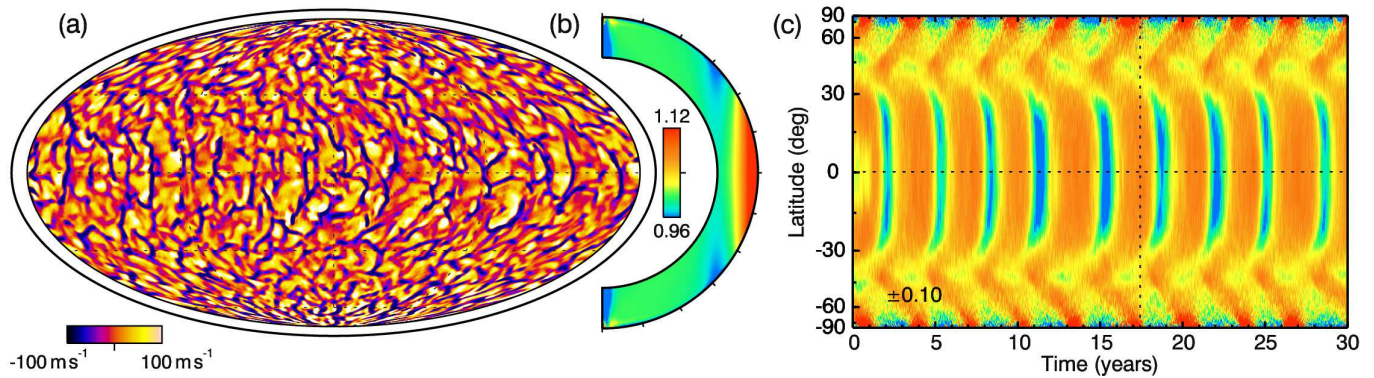} \figcaption{Convective
       patterns and differential rotation. (a) Snapshot of the horizontal convective patterns
       arising in the radial velocity $v_r$ at $\Rsun{0.95}$ shown in Mollweide projection, at the
       time corresponding to the vertical dashed line in (c). This reveals the larger-scale
       convection at low latitudes and the smaller-scales at higher latitudes, with downflows dark
       and upflows in lighter tones. (b) Time and azimuthally-averaged angular velocity
       $\avg{\avgO}$ (double brackets indicating dual averages), showing a fast equator in red and
       slower high-latitudes in blue. (c) A time-latitude diagram of azimuthally-averaged
       $\avg{\Delta\Omega}=\avgO-\avg{\avgO}$ in cylindrical projection, elucidating the propagation
       of equatorial and polar branches of a torsional oscillation arising from strong Lorentz-force
       feedback. The color indicates enhanced differential rotation in red and periods of slower
       rotation in blue, with variations of up to $\pm 10$\% of the bulk rotation rate.\label{fig2}}
   \end{center}
\end{figure*}

Inspired by these recent ASH and EULAG results, we have attempted to splice the two together by
incorporating SLD into ASH with the express goal of achieving a low effective $\Prt$ and $\Prm$
dynamo. Thus an attempt is made to better mimic the low Prandtl number solar setting, while keeping
the eddy-diffusive approximation for entropy mixing and treating the reconnection of small-scale
magnetic field as diffusive. This effort minimizes the effects of viscosity, and so extends the
inertial range as far as possible for a given resolution. Thus SLD permits more scales to be
captured before entering the dissipation range, allowing more scale separation between the larger
magnetic and and smaller kinetic scales participating in the low $\Prm$ dynamo \citep{ponty05,
  schekochihin07, brandenburg09}. Subsequently, the kinetic helicity is also greater at small scales
than otherwise would be achieved, which has been shown to have a large influence on the dynamo
efficiency \citep{malyshkin10}. Indeed, with this newly implemented diffusion minimization scheme,
we have happened upon a solution that possesses four fundamental features of the solar dynamo: a
regular magnetic energy cycle period, and an orderly magnetic polarity cycle of $\tau_C=6.2$~years,
equatorward propagation of magnetic features, and poleward migration of oppositely signed
flux. Furthermore this equilibrium is punctuated by an interval of relative quiescence, after which
the cycle is recovered. In keeping with the ASH nomenclature for cases as in \citep{brown10,
  brown11, nelson13a}, this dynamo solution has been called D3S.

Figure \ref{fig1} illustrates the morphology of the toroidal fields in space and time. The presence
of large-scale and azimuthally-connected structures is evident in Figures \ref{fig1}(a, d). Such
toroidal structures have been dubbed wreaths \citep{brown10}. In D3S, there are two
counter-polarized, lower-latitude wreaths that form near the point where the tangent cylinder
intersects a given spherical shell. This point is also where the peak in the latitudinal gradient of
the differential rotation exists for much of a magnetic energy cycle. There are also polar caps of
magnetism of the opposite sense of those at lower latitudes. These caps serve to moderate the polar
differential rotation, which would otherwise tend to accelerate and hence establish fast polar
vortices. The average structure of the wreaths and caps is apparent in Figure \ref{fig1}(b), which
is averaged over a single energy cycle or 3.1~years. The wreaths appear rooted at the base of the
convection zone, whereas the caps have the bulk of their energy in the lower convection zone above
its base. This is somewhat deceptive as the wreaths are initially generated higher in the convection
zone, while the wreath generation mechanism (primarily the $\Omega$-effect) migrates equatorward and
toward the base of the convection zone over the course of the cycle. The wreaths obtain their
greatest amplitude at the base of the convection zone and thus appear seated there.

Figure \ref{fig2}(a) shows a typical convective pattern during a cycle, with elongated and
north-south aligned flows at low latitudes and smaller scales at higher latitudes. In aggregate, the
spatial structure and flow directions along these cells produce strong Reynolds stresses acting to
accelerate the equator and slow the poles. In concert with a thermal wind, such stresses serve to
rebuild and maintain the differential rotation during each cycle. While the variable nature of the
convective patterns over a cycle is not shown, it is an important piece of the story. Indeed, the
magnetic fields disrupt the alignment and correlations of these cells through Lorentz forces.
Particularly, as the field gathers strength during a cycle, the strong azimuthally-connected
toroidal fields tend to create a thermal shadow that weakens the thermal driving of the equatorial
cells. Thus their angular momentum transport is also diminished, which explains why the differential
rotation seen in Figure \ref{fig2}(b) cannot be fully maintained during the cycle. This is captured
in the ebb and flow of the kinetic energy contained in the fluctuating velocity field, which here
vary by about 50\%. Such a mechanism is in keeping with the impacts of strong toroidal fields in the
convection zone suggested by \citet{parker87}. Moreover, strong nonlinear Lorentz force feedbacks
have been seen in other convective dynamo simulations as well \citep{brown11}, and they have been
theoretically realized for quite some time in mean-field theory \citep[e.g.,][]{malkus75}.

\section{Cycle Periods} \label{sec:periods}

There are a large set of possible and often interlinked time scales that could be relevant to the
processes setting the pace of the cyclical dynamo established in D3S. For instance, there are
resistive time scales that depend upon the length scale chosen. One such time scale is the resistive
decay of the poloidal field at the upper boundary as it propagates from the tangent cylinder to the
equator, which would imply that the length scale is $\ell = r_2 \Delta\theta$ and so $\tau_{\eta} =
\ell^2/\eta_2 \approx 6.7$~years and is close to the polarity cycle period, where the subscript two
denotes the value of a quantity at the outer boundary of the simulation. However, this is likely not
dynamically dominant as the the polarity reversal occurs in half that time. The same is true of the
diffusion time across the convection zone, being $4.6$~years. Since the cycle is likely not
resistively controlled it must be the interplay of dynamical processes. Another mechanism to
consider is the cycle time related to flux transport by the meridional flow, then the transit time
of a magnetic element along its circuits could be relevant. In D3S, the mean meridional flow is
anti-symmetric about the equator and has two cells, with a polar branch and a lower latitude cell
that are split by the tangent cylinder. The circulation time of the polar branch is about 0.7~years,
whereas that of the equatorial cell is about a year. So it is also unlikely that the meridional flow
is setting the cycle period.

\begin{figure}[t!]
   \begin{center}
     \includegraphics[width=0.45\textwidth]{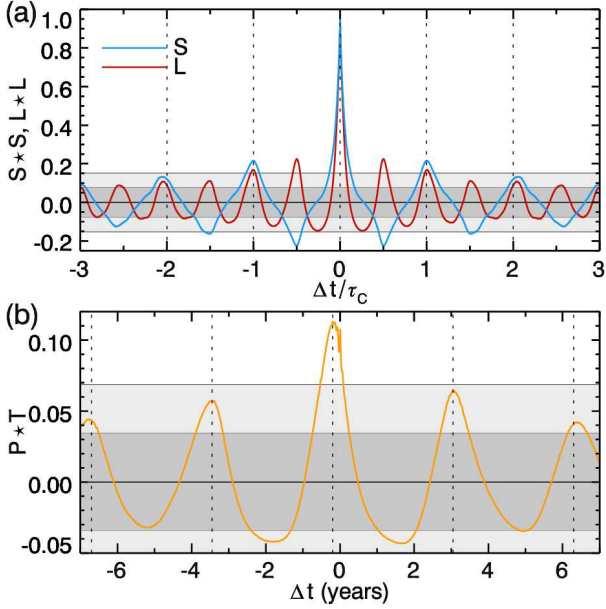} 
     \caption[Auto and cross-correlation of Lorentz-force and toroidal field production by
       mean-shear]{(a) Volume-averaged temporal auto-correlation of toroidal magnetic energy
       generation by mean shear ($\mathrm{S} = \lambda \avg{\vB_{P}} \cnabla \avgO$, blue curve) and
       the same for the mean Lorentz force impacting the mean angular velocity ($\mathrm{L}_{\phi}$,
       red curve) plotted against temporal lags $\Delta \mathrm{t}$ normalized by the polarity cycle
       period $\tau_C=6.2$~years. Confidence intervals are shown as shaded gray regions, with the
       67\% interval in darker gray and 95\% in lighter gray. (b) Cross-correlation of the mean
       poloidal energy production ($\mathrm{P} = \mathbf{B}_P\cnabla\cross\mathcal{E}'_{\phi}$)
       through the fluctuating EMF and the toroidal magnetic energy production due to the mean shear
       ($\mathrm{T} = \avgBp S$), showing the nonlinear dynamo wave character of the
       solution. \label{fig3}}
   \end{center}
\end{figure}

The dynamical coupling of azimuthally-averaged magnetic fields $\avg{\vB}$ and the mean angular
velocity $\avgO$ (Figure \ref{fig2}(b)) plays a crucial role in regulating the cycle, though it
alone cannot be the sole actor as is well known from Cowling's anti-dynamo theorem. The significant
anti-correlation of $\avgBp$ and angular velocity variations $\avg{\Delta\Omega}$ during reversals
becomes apparent when comparing Figures \ref{fig1}(c) and \ref{fig2}(c), revealing the strong
nonlinear coupling of the magnetic field and the large-scale flows. The dynamics that couples these
two fields is the toroidal field generation through the mean shear ($\mathrm{S} = \lambda
\avg{\vB_{P}} \cnabla \avgO$, with $\avg{\vB_{P}}$ the mean poloidal field) and the mean azimuthal
Lorentz-force ($\Lp = \boldsymbol{\hat{\phi}\cdot} \avg{\vJ} \cross \avg{\vB}$), which acts to
decrease $\avgO$. The auto-correlation of each of these components of the MHD system reveals that
$\Lp$ varies with a period corresponding to the magnetic energy cycle, whereas $\mathrm{S}$ varies
on the polarity cycle period (Figure \ref{fig3}).  It also shows the high degree of temporal
self-similarity between cycles, with the auto-correlation of both quantities remaining significant
with 95\% confidence for a single polarity cycle and with 67\% confidence for three such cycles.

Appealing to Figure \ref{fig1}(c), it is evident that $\vB$ exhibits a high degree of
spatial and temporal self-similarity, though with reversing polarity. Thus the period apparent in
the auto-correlation for $\Lp$ might be expected. Furthermore, if we simply let $\avg{\vB} \approx
\vB_0(r,\theta) \exp(i\omega_C t)$, the Lorentz forces could be characterized very roughly as $\Lp
\propto {\Lp}_{, 0} \exp(i \omega_L t) \sim \vB_0\cdot\vB_0/\ell \exp(2 i \omega_C t)$, with cycle
frequency $\omega_C = 2\pi/\tau_C$ and some length scale $\ell$. Hence, the magnetic energy or
Lorentz cycle frequency $\omega_L = 2\pi/\tau_{L}$ implies that $2 \tau_L = \tau_C$.  What is
potentially more curious is that $\mathrm{S}$ varies on the cycle period. While Figure \ref{fig2}(c)
might suggest a reversal in the solar-like character of the differential rotation. This in fact does
not occur. Rather, the shear is significantly weakened but maintains the positive latitudinal
gradient that sustains the toroidal magnetic field, which renders the sign of $\grad\Omega$
independent of time. Therefore, the polarity reversals in $\avg{\vB_P}$ require that $\mathrm{S}$
vary with the polarity cycle period $\tau_C$.

\begin{figure}[t!]
   \begin{center}
   \includegraphics[width=0.45\textwidth]{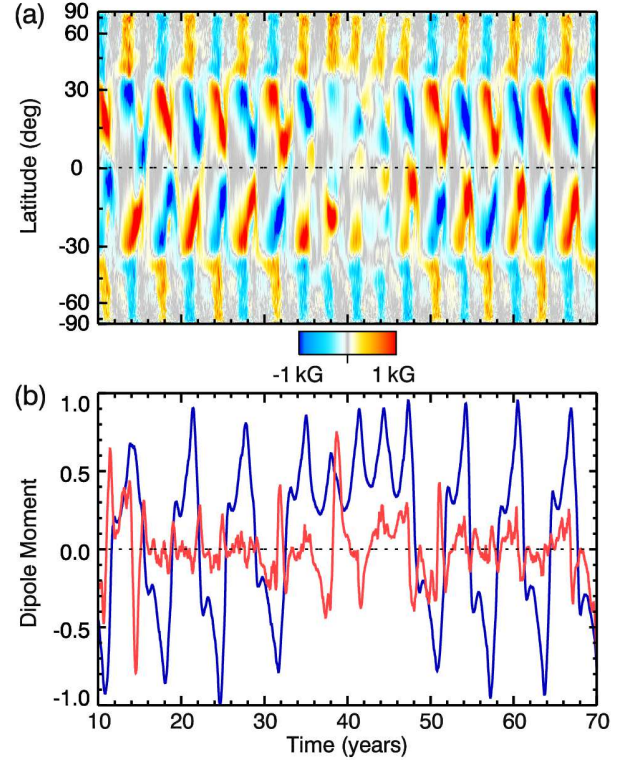} \figcaption{An interval of
     magnetic quiescence. (a) Time-latitude diagram of $\avgBp$ at $\Rsun{0.95}$ in cylindrical
     projection, picturing the loss and reappearance of cyclical polarity reversals as well as the
     lower amplitude of the wreaths. Strong positive toroidal field is shown as red, negative in
     blue. (b) Normalized magnetic dipole moment (red) and the quarupolar moment (blue). The
     quadrupole moment peaks near reversals, indicating its importance. \label{fig4}}
   \end{center}
\end{figure}

\section{Equatorward Propagation} \label{sec:propagate}

As with ASH and EULAG, simulations in spherical segments that employ the Pencil code also obtain
regular cyclical magnetic behavior. Some of these polarity reversing solutions exhibiting
equatorward propagating magnetic features \citep{kapyla12}, magnetic flux ejection
\citep{warnecke12}, and 33-year magnetic polarity cycles \citep{warnecke13}. Currently, however, the
mechanism for the equatorward propagation of the magnetic structures in those simulations remains
unclear. Perhaps the mechanism is similar to that seen here.

The equatorward propagation of magnetic features observed in this case, as in Figures \ref{fig1}(c)
and \ref{fig4}(a), arises through two mechanisms. The first process is the nonlinear feedback of the
Lorentz force that acts to quench the differential rotation, disrupting the convective patterns and
the shear-sustaining Reynolds stresses they possess. Since the latitudinal shear serves to build and
maintain the magnetic wreaths, the latitude of peak magnetic energy corresponds to that of the
greatest shear. So the region with available shear moves progressively closer to the equator as the
Lorentz forces of the wreaths locally weaken the shear. Such a mechanism explains the periodic
modifications of the differential rotation seen in Figure \ref{fig2}(c). However, it does not
explain how this propagation is initiated and sustained, as one might instead expect an equilibrium
to be established with the magnetic energy generation balancing the production of shear and which is
further moderated by cross-equatorially magnetic flux cancellation as the distance between the
wreaths declines.

There are two possibilities for the second mechanism that promotes the equatorward propagation of
toroidal magnetic field structures. If we may consider the dynamo action in this case as a dynamo
wave, the velocity of the dynamo wave propagation is sensitive to the gradients in the angular
velocity and the kinetic helicity in the context of an $\alpha\Omega$ dynamo
\citep[e.g.,][]{parker55,yoshimura75}. A simple analysis indicates that near and poleward of the
edge of the low-latitude wreaths the sign of the Parker-Yoshimura mechanism is correct to push the
dynamo wave toward the equator, but the effect is marginal elsewhere. The second possibility is that
the spatial and temporal offsets between the fluctuating EMF and the mean-shear production of
toroidal field leads to a nonlinear inducement to move equatorward. This mechanism relies on the
concurrent movement of the turbulent production of the poloidal field that continues to destroy
gradients in angular velocity through the production of toroidal magnetic through the action of the
differential rotation on the renewed poloidal field. Nonetheless, the wreaths eventually lose their
azimuthal coherence because of cross-equatorial flux cancellation and the lack of sufficient
differential rotation to sustain them, which leads to a rapid dissemination of the remaining flux by
the convection. This is evident in Figure \ref{fig1}(a), where at the end of each cycle the wreaths
converge on the equator and their resulting destruction leads to the poleward advection of
field. This advected field is of the opposite sense of the previous cycle's polar cap and, being of
greater amplitude compared to the remaining polar field, establishes the sense of the subsequent
cycle's polar field. Furthermore, in D3S, as a cycle progresses the centroid for the greatest dynamo
action propagates both equatorward and downward in radius, as might be deduced from the successful
reversals visible in Figure \ref{fig4}(b). Though, it is more evident in a time-radius diagram.
Hence, the equatorial migration begun at the surface makes its way deeper into the domain as the
cycle progresses.

\section{Grand Minima} \label{sec:intermit}

As with some other dynamo simulations \citep[e.g.,][]{brown11,augustson13}, there is also long-term
modulation in case D3S. Figure \ref{fig4} shows an interval of about 20 years where the polarity
cycles are lost, though the magnetic energy cycles resulting from the nonlinear interaction of the
differential rotation and the Lorentz force remains. During this period, the magnetic energy in the
domain is about 25\% lower, whereas the energy in the volume encompassed by the lower-latitudes is
decreased by 60\%. However, both the spatial and temporal coherency of the cycles are recovered
after this interval and persist for the last 40~years of the 100~year-long simulation.  Prior to
entering this quiescent period, there was an atypical cycle with only the northern hemisphere
exhibiting equatorward propagation. This cycle also exhibits a prominent loss of the equatorial
anti-symmetry in its magnetic polarity. The subsequent four energy cycles do not reverse their
polarity, which is especially evident in the polar regions, whereas the lower latitudes do seem to
attempt such reversals.

\section{Conclusions} \label{sec:conclude}

The simulation presented here is the first to self-consistently exhibit four prominent aspects of
solar magnetism: regular magnetic energy cycles during which the magnetic polarity reverses, akin to
the sunspot cycle; magnetic polarity cycles with a period of 6.2~years, where the orientation of the
dipole moment returns to that of the initial condition; the equatorward migration of toroidal field
structures during these cycles; and quiescence after which the previous polarity cycle is
recovered. Furthermore, this simulation may capture some aspects of the influence of a layer of
near-surface shear, with a weak negative gradient in $\avgO$ within the upper 10\% of the
computation domain (3\% by solar radius). The magnetic energy cycles with the time scale $\tau_C/2$
arise through the nonlinear interaction of the differential rotation and the Lorentz force. We find
that the nonlinear feedback of the Lorentz force on the differential rotation significantly reduces
its role in the generation of toroidal magnetic energy. The magnetic fields further quench the
differential rotation by impacting the convective angular momentum transport during the
reversal. Furthermore, despite the nonlinearity of the case, there is an eligible influence of a
dynamo wave in the fluctuating production of poloidal magnetic field linked to the shear-produced
toroidal field. The mechanisms producing the equatorward propagation of the toroidal fields have
been identified, with the location of the greatest latitudinal shear at a given point in the cycle
and the weak negative radial shear both playing a role. This simulation has also exhibited
long-lasting minimum, loosely similar to the Maunder minimum. Indeed, there is an interval covering
20\% of the cycles during which the polarity does not reverse and the magnetic energy is
substantially reduced. Despite rotating three times faster than the Sun and parameterizing large
portions of its vast range of spatio-temporal scales, some of the features of the dynamo that may be
active within the Sun's interior have been realized in this global-scale ASH simulation.

\section*{Acknowledgments}

The authors thank Nicholas Featherstone, Brad Hindman, Mark Rast, Matthias Rempel, and Regner
Trampedach, for helpful and insightful conversations. This research is primarily supported by NASA
through the Heliophysics Theory Program grant NNX11AJ36G, with additional support for Augustson
through the NASA NESSF program by award NNX10AM74H. The computations were primarily carried out on
Pleiades at NASA Ames with SMD grants g26133 and s0943, and also used XSEDE resources for
analysis. This work also utilized the Janus supercomputer, which is supported by the NSF award
CNS-0821794 and the University of Colorado Boulder.

\bibliography{apj-jour,d3let}

\end{document}